\documentclass[prl,aps,twocolumn,showpacs,amsmath,amssymb,floatfix]{revtex4}

\usepackage{graphicx}% Include figure files
\usepackage{dcolumn}% Align table columns on decimal point
\usepackage{bm}% bold math
\usepackage{epsfig}
\usepackage{amsmath}

\newcommand{\be}{\begin{equation}}
\newcommand{\ee}{\end{equation}}
\newcommand{\beq}{\begin{eqnarray}}
\newcommand{\enq}{\end{eqnarray}}
\newcommand{\ua}{\uparrow}
\newcommand{\da}{\downarrow}
\newcommand{\eqname}[1]{\label{eq:#1}}
\newcommand{\eq}[1]{(\ref{eq:#1})}

\begin{document}

%\preprint{APS/123-QED}

\title{Spin fluctuations, susceptibility and the dipole oscillation of a nearly ferromagnetic Fermi gas} 

\author{Alessio Recati and Sandro Stringari}
\affiliation{Dipartimento di Fisica, Universit\`a di Trento and INO-CNR BEC Center, I-38123 Povo, Italy}
 
\begin{abstract}
We discuss the spin fluctuations and the role played by the magnetic susceptibility  in an atomic Fermi gas interacting with positive scattering length. Both thermal and quantum fluctuations are considered. Using a sum rule approach and recent {\it
ab initio}  Monte Carlo results for the magnetic susceptibility of uniform matter we provide explicit predictions for the frequency of the spin dipole oscillation of a gas trapped by a harmonic potential and discuss the deviations from the behaviour of an ideal gas when the system approaches the ferromagnetic transition. The role of the Landau's parameters in the characterization of the magnetic properties is also discussed.
\end{abstract} 

\pacs{03.75.Kk,0375.Ss,67.25.D-}

\maketitle

The recent experiment on itinerant ferromagnetism reported in Ref.~\cite{itinferr} has stimulated a novel debate on the magnetic properties of interacting Fermi gases \cite{duine,conduitsimons1,conduitsimons2,debate,Hui,conduitaltman,demler}. In this experiment it was shown that, by increasing the value of the positive scattering length $a$, the system exhibits  characteristic features of the Stoner instability where the gas enters a new phase of ferromagnetic type. The experiment was performed on a metastable state of a Fermi gas. Actually, due to the attractive nature of the interaction, there exists a 2-body bound state for positive values of the scattering length and the ground state of the system is a Bose gas of molecules (see, e.g., \cite{RMPtrento} and references therein). Indeed, three-body collisions, which convert atoms into molecules, provide one of the signatures of the transition to a magnetic phase, since they become more relevant by increasing $a$ in the normal state, but are less efficient in the magnetic phase due to the formation of spatial domains of non interacting atoms. They give rise to a maximum in the atomic losses. Other signatures of the phase transition were a maximum in the cloud size and a minimum in the kinetic energy \cite{itinferr}. This experiment did not however measure directly any magnetic property of the system. In particular no evidence for magnetic domains was found above the transition point. Recent Monte Carlo simulations \cite{giorgini,trivedi} have shown that the ferromagnetic transition of an interacting Fermi gas takes place at the value $k_Fa\sim 0.8-0.9$, where $k_F=(3 \pi^2 n)^{1/3}$ is the Fermi momentum and $n$ the density of the sample. This value is significantly smaller than the experimental value found at the lowest temperatures. This discrepancy, together with the absence of evidence for magnetic domains,  suggests that the physical behaviour of the system is far from being understood. In particular the occurrence of 3-body losses \cite{conduitsimons2,conduitaltman} and the competition between the tendency of the system to become ferromagnetic and to form pairs   \cite{demler} are likely at the origin of new physical effects.  

One of the most relevant thermodynamic quantities characterising the transition to the ferromagnetic phase is the magnetic susceptibility $\chi=(V \partial^2 A/\partial (N_\ua-N_\da)^2)^{-1}$, where $A$ is the free energy, $V$ the volume of the sample and $N_\uparrow$ and $N_\downarrow$ are the number of fermions in the two spin-states. In general the interaction will increase the susceptibility of the gas (see, e.g., \cite{Pathria}) which, in a second order phase transition, should diverge at the transition. The actual order of the transition is sensitive to temperature, disorder and polarisation (see, e.g., \cite{duine,conduitsimons1,MnSi,BelitzVojta} and reference therein) \cite{noteOrder}. 

At finite temperature the spin susceptibility can be related to the measurement of the thermal spin fluctuations:
\be
\frac{\Delta (N_\uparrow-N_\downarrow)^2}{N}= k_{B}T \frac{\chi(T)}{n}
\eqname{Tfluc}
\ee
where $N=N_\uparrow+N_\downarrow$ and $n$ is the density of the sample. 
The measurement of the spin fluctuations, and through \eq{Tfluc} of $\chi$, seems to be now accessible, as it has been shown by the very recent shot noise measurements of the density fluctuations in an ideal Fermi gas \cite{ETH,mit}. In such experiments the fluctuations are measured in a small sub-volume of the whole atomic cloud, where the system can be considered  uniform \cite{S}. 

At low temperature the applicability of \eq{Tfluc} becomes less and less efficient because of the  suppression
 caused by the thermal factor and the emergence of quantum fluctuations which exhibit a weaker dependence on $N$, but become soon important as $T\to 0$. The quantum fluctuations of the number of particles have been investigated in \cite{AstraLev}. The result for the spin fluctuations, to the leading order in $N$, can be written as
\be
\frac{\Delta (N_\uparrow-N_\downarrow)^2}{N}=2 \alpha \left(\frac{12}{\pi^4 N}\right)^{1/3}\ln(C N^{1/3}),
\eqname{qfluc}
\ee 
where the parameter $\alpha$ is fixed by the low $q$ behavior of the static structure factor, according to $S^a(q\to 0)=\alpha q/k_F$ and $C$ is a cut-off constant determined by the short range behaviour of $S^a(q)$. 
%Note that the density static structure factor is accessible for ultra-cold gases \cite{expSq}. 
In the case of the ideal Fermi gas one finds $\alpha=3/2^{7/3}$, $C\sim 10.45$. Thus the comparison with Eq. \eq{Tfluc} using the $T=0$ ideal Fermi gas susceptibility value $\chi_0=3n/2\epsilon_F$, with $\epsilon_F$ the Fermi energy, shows that quantum fluctuations become dominant for temperatures  $T<T_F/N^{1/3}(0.92+0.12 \ln N)$, where $T_F=\epsilon_F/k_B$ is the Fermi temperature. Thus, taking for example, $T\simeq 0.2 T_F$ quantum fluctuations become important already for $N\le 10^3$. 

For interacting Fermi gases the value of $\chi$ and $\alpha$ can be calculated employing Landau theory of Fermi liquids. Moreover the coefficient $\alpha$ is bounded by the spin susceptibility through the sum rule inequality $\alpha\le \sqrt{\epsilon_F\chi(T=0)/2 n}$. Thus, an important question is whether and how also $\alpha$ -- and consequently the $T=0$ quantum fluctuations -- diverges at the transition point identified by the divergence in $\chi$.

In Landau theory of Fermi liquids the susceptibility is given by 
\begin{equation}
{\chi_0\over \chi}=(1+F_0^a){m\over m^*}\; ,
\label{chi0chi}
\end{equation}
where $m^*=m(1+F_1^s/3)>m$ is the effective mass, and $F_l^s$, $F_l^a$ are the $l$-th angular momentum symmetric (density) and antisymmetric (spin) Landau parameters, respectively. The susceptibility and thus the thermal fluctuations \eq{Tfluc} present a divergence for $F_0^a=-1$ \cite{he3}.

The parameter $\alpha$ can also be expressed in terms of the Landau parameters. Indeed, in general, $S^a(q)=-\pi^{-1}\int Im\chi(q,\omega)$ where $\chi(q,\omega)$  is the linear response function determined by the complete quasi-particle scattering amplitude \cite{Pines}. It is well known that the linear response function presents a pole for $\omega=0$ at $F_0^a=-1$ \cite{Pines}. By taking into account only the $l=0$ Landau parameters, $\alpha$ reads
\be
\alpha=
\frac{3}{2^{4/3}}\int_0^1 \frac{\lambda d\lambda}{\left(1-F_0^a(1-\frac{\lambda}{2}\ln\frac{1+\lambda}{1-\lambda})\right)^2+\left(\frac{\lambda}{2} F_0^a\right)^2}
\ee 
and, thus it  also diverges for $F_0^a=-1$, but just logarithmically, $\alpha\propto \ln(1/(1+F_0^a))$. This means that quantum fluctuations increase more slowly than thermal fluctuations as one approaches the critical point of the ferromagnetic transition. 

\begin{figure}%[ptb]
\begin{center}
\includegraphics[height=6cm]{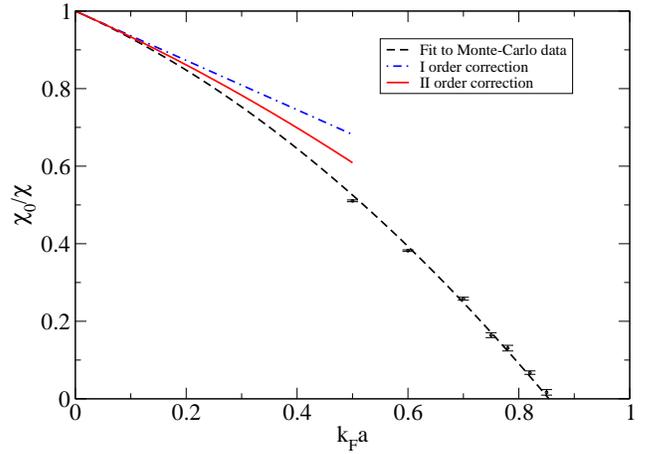}
\caption{Inverse spin susceptibility for an homogeneous Fermi gas as a function of the interaction parameter $k_F a$. Points: Monte-Carlo data from \cite{giorgini}. Dashed (green) line: fit to the Monte-Carlo data. The dot-dashed (blue) and the continuous (red) lines are the first and second order expansion, respectively (see Eq. \eq{expansion}). }
\label{fig:susc}
\end{center}
\end{figure}

When the interaction is small the Landau parameters can be determined using perturbation theory \cite{Lev} and, to second order in $k_F a$, they read \cite{forward}
\begin{eqnarray}
F_0^s=\frac{2}{\pi}(k_F a)+\frac{8}{3\pi^2}(2+\ln 2) (k_F a)^2,
\eqname{F0s}\\
F_1^{s} = \frac{8}{5\pi^2}(7\ln 2-1)(k_Fa)^2,
\eqname{F1s}\\
F_0^a=-\frac{2}{\pi}(k_F a)-\frac{8}{3\pi^2}(1-\ln 2)(k_F a)^2,
\eqname{F0a}\\
F_1^{a} = -\frac{8}{5\pi^2} (2+ \ln 2) (k_Fa)^2,
\eqname{F1a}
\end{eqnarray}
so that the inverse susceptibilty, to the same order, takes the form
\be
\frac{\chi_0}{\chi}=\left[1-\frac{2}{\pi} k_Fa -\frac{16 \;(2+ \ln 2)}{15\pi^2}(k_Fa)^2\right].
\eqname{expansion}
\ee
To the same order one can also calculate the inverse compressibility  
 of the gas: $\kappa^0/\kappa=(1+F_0^s)m/m^*$, where $\kappa_0= 3/2 n \epsilon_F$ the ideal Fermi gas compressibility. For higher values of $k_Fa$
the susceptibility can be calculated numerically using quantum Monte-Carlo  techniques \cite{giorgini}. The results for $\chi^{-1}$ as a function of the dimensionless parameter $k_Fa$ are shown in Fig. \ref{fig:susc} where the Monte Carlo predictions are compared  with the small $k_Fa$ expansion \eq{expansion}. The figure shows that, a ferromagnetic instability appears around $(k_F a)_c\simeq 0.83$. The first order expansion of $\chi^{-1}$, extrapolated to the critical point, gives the well known Stoner result $(k_F a)_c=\pi/2\simeq 1.57$, while the second order expansion would give $(k_F a)_c\simeq 1.05$.

Aside from the fluctuations, another way to investigate the magnetic properties of Fermi gases is the study of their dynamic behaviour as a response to a spin perturbation. In this letter we analyse the zero-temperature spin dipole oscillation of a two component Fermi gas in the presence of an harmonic trap $V({\bf r})=1/2 m (\omega_x^2 x^2+\omega_y^2 y^2+\omega_z^2 z^2)$. The dipole mode for a weakly interacting Fermi gas has been already studied in \cite{Vichi} and experimental results were reported in \cite{JinSD}. In this letter we estimated the frequency of this oscillation using a sum rule approach \cite{sumrules} based on the calculation of the ratio
\begin{equation}
\hbar^2\omega^2_{SD} = {m_1\over m_{-1}}
\label{1-1}
\end{equation}
where $m_k=\sum_n|\langle 0|D|n\rangle|^2(E_n-E_0)^k$ are the moment of the strength distribution function relative to the spin dipole operator $D=\sum_i(z_{i\uparrow}-z_{i\downarrow})$. The ratio (\ref{1-1}) actually provides an upper bound to the frequency of the lowest energy state excited by the  operator $D$. Differently from uniform matter, where a negative value of the spin parameter $F_0^a$  gives rise  to Landau damping, in  the presence of harmonic trapping one expects a discretized collective  mode with frequency smaller than the free oscillator value.

The energy weighted sum rule entering the numerator of (\ref{1-1}) is easily calculated in terms of a double commutator:
\begin{equation}
m_1= \frac{1}{2}\langle 0 |[D,[H,D]]|0\rangle = N{\hbar^2\over 2 m}
\label{f}
\end{equation}
where we have used the fact that only the kinetic energy operator gives a contribution to the commutator $[H,D]$. The inverse energy weighted sum rule is instead  related to the spin dipole  polarizability and  can be calculated, at zero temperature, by minimizing the total energy of the system in the presence of an external static coupling of the form $-\lambda D$. In the local density approximation the energy can be written as:
\begin{equation}
E= \int d{\bf r}[\epsilon(n_\uparrow({\bf r}),n_\downarrow({\bf r})) -\lambda z(n_\uparrow({\bf r})-n_\downarrow({\bf r}))]
\label{E}
\end{equation}
where $\epsilon(n_\uparrow,n_\downarrow)$ is the energy density of  uniform matter. By expanding $\epsilon(n_\uparrow,n_\downarrow)$  up to quadratic terms in $n_\uparrow-n_\downarrow$,  minimization of $E$ yields the result $
n_\uparrow-n_\downarrow= \lambda z \ chi(n)
%\label{pol}
$
for the polarization induced by the external field where $
\chi^{-1}(n)=\partial^2 \epsilon/\partial (n_\ua-n_\da)^2$ is the zero temperature inverse magnetic susceptibility of uniform matter calculated at the local value of the density. The calculation of the induced dipole spin moment then yields $m_{-1}= 1/2\int d{\bf r} z^2 \chi(n)$ and hence the result
\begin{equation}
\omega^2_{SD}={N\over m\int d{\bf r} z^2 \chi(n)}  
\label{chichi0}
\end{equation}
for the dipole frequency. Eq. (\ref{chichi0}) shows explicitly that an increase of the magnetic susceptibility  will result in a decrease of $\omega_{SD}$ \cite{giantres}. The density profile, $n({\bf r})$, entering the integral (\ref{chichi0}) should be calculated in the local density approximation using the equilibrium condition
$\mu(n) + V_{ho}=\mu_0$ where $\mu(n)= \partial \epsilon(n)/\partial n$ is the chemical potential and $\mu_0$ is fixed by the normalization condition. 
One can easily check that, using the ideal gas expression for the magnetic susceptibility, Eq. (\ref{chichi0}) yields the expected result $\omega_{SD}=\omega_z$  where $\omega_z$ is the trapping oscillator frequency in the $z$-th direction. Inclusion of interaction effects modifies the value of the spin dipole frequency both through the change of the magnetic susceptibility $\chi$ and the change of the density profile fixed by the equation of state $\mu(n)$. At the first order in the scattering lenght we get 
\be 
\omega_{SD}=\omega_z\left(1-\frac{128\sqrt{2}\;(3N)^{1/6}}{35\pi^2}\frac{a}{a_{ho}}\right),
\label{wIorder}
\ee
where $a_{ho}$ is the geometrical average of harmonic oscillator lengths.
Result (\ref{wIorder}) coincides with the prediction made in \cite{Vichi} employing a scaling ansatz. Such a coincidence is not surprising since one can easily show that, in the limit of the ideal Fermi gas, the polarization induced by the static dipole field coincides with the scaling form $n_\uparrow-n_\downarrow = (\lambda/m\omega^2_z) \partial_zn$. According to perturbation theory result (\ref{wIorder}) then corresponds to an exact result for the spin dipole frequency calculated to first order in $a$.
\begin{figure}%[ptb]
\begin{center}
\includegraphics[height=6cm]{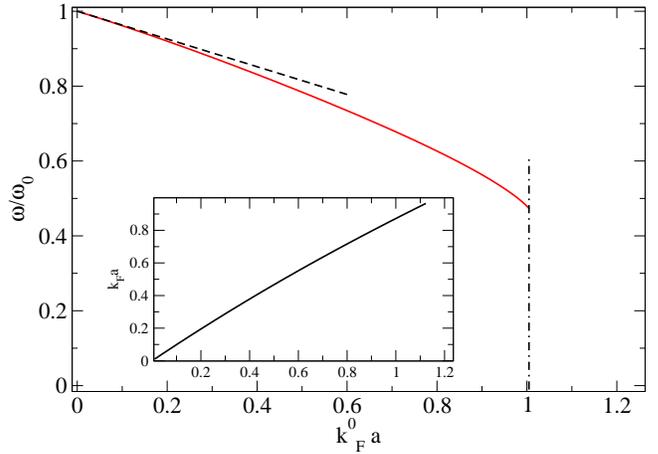}
\caption{Main panel: Spin dipole frequency as a function of the interaction parameter $k_F^0 a$ (see text). The dashed line is the first order expansion Eq. (\ref{wIorder}). Inset: the value of the interaction parameter $k_F a$ in the trap as a function of $k_F^0 a$. $k_F$ is the Fermi momentum of the interacting cloud calculated at the centre of the trap.}
\label{fig:freqSD}
\end{center}
\end{figure}

In Fig. \ref{fig:freqSD} we report the value of the spin dipole frequency (\ref{chichi0}) calculated using the Monte-Carlo results for $\mu(n)$ and $\chi(n)$ of uniform matter \cite{giorgini} and using the local density approximation to account for the inhomogeneity caused by the external potential. The results are reported as a function of $k_F^0a$, where $k_F^0=(24 N)^{1/6}/a_{ho}$ is the Fermi momentum of an ideal Fermi gas calculated in the centre of the trap, $a_{ho}$ is the the geometrical average of the harmonic oscillator lengths. In the inset we report the value of the interaction parameter $k_F a$ as a function of $k_F^0 a$. The instability, which appears initially in the centre of the trap at around $k_F^0 a\simeq 1$ corresponds to the (Monte-Carlo) bulk value $k_F a\simeq 0.83$ calculated in the centre of the trap. The dipole frequency is seen to be significantly affected by the interaction even for values of $k_F^0a$ well below the ferromagnetic transition. This result is promising in view of the possibility of checking experimentally our understanding of the magnetic properties of the Fermi gas in the regime where the expansion in powers of $k_F^0a$ should be universal (i.e. independent of the actual form of the two-body potential) and where the life time of the sample is large enough  \cite{itinferr}. At the transition point the dipole frequency (\ref{1-1}) does not vanish since the divergence of the magnetic susceptibility concerns only the behaviour of the gas in the centre of the trap where the integral (\ref{chichi0}) is suppressed by the factor $z^2$.

It is worth noting that the result (\ref{chichi0}) provides an upper bound to the lowest spin dipole frequency.  An even lower bound can be obtained by noticing that, according to Landau theory of Fermi liquids, the $f$-sum rule in the spin channel is not exhausted by the low-lying spin excitations. Indeed the low-energy quasi-particle excitations give a contribution to the $f$-sum rule proportional to the combination $m/m^*(1+1/3F_1^{s(a)})$. Thus, while in the density channel the sum rule is not affected by the interaction, the spin dipole $f$-sum rule is instead reduced by the interaction since $F_1^s>F_1^a$, the remaining part of the $f$-sum rule being provided by multi-pair excitations located at higher energy \cite{FrancoSS}. The renormalized value of the  $m_1$ sum rule up to the second order in $k_F a$ can be explicitly determined using the expressions \eq{F1s} and \eq{F1a} for the Landau parameters. We find:
\begin{equation}
m_1 = \!\!{\hbar^2\over m}\int d{\bf r}n({\bf r})\!\left[1 -\frac{8 (3 \pi^2)^{2/3}}{15\pi^2} (1+ 8\ln 2) (k_F({\bf r})a)^2\right].
\label{m1R}
\end{equation}
The explicit calculation of the integral (\ref{m1R}) yields a reduction of the spin dipole frequency of less than 10 per cent at $k_F^0a=0.5$ with respect to the prediction of (\ref{chichi0}). 

Let us conclude by pointing out that an open issue concerns the effects of collisions on the damping of the spin
dipole mode and in particular its dependence on temperature and on the value of the
scattering length expecially close to the transition \cite{DuineDrag}.

\paragraph{Aknowledgment.}

We thank G. Bertaina for the Monte-Carlo data and F. Piazza for careful reading and useful comments on the manuscript. We also acknowledge useful discussions with W. Ketterle, L. P. Pitaevskii and W. Zwerger. This research has been supported by EuroQUAM Fermix program and by MIUR PRIN 2007.


\begin{thebibliography}{99}
\bibitem{noteOrder}
In particular for a clean and not polarised system the transition is ``weakly'' first order till a certain temperature, above which is second order. In the most recent Monte-Carlo calculation \cite{giorgini} at $T=0$ for cold gases is very hard to distinguish whether the transition is of first or of the second order and a divergent behaviour of $\chi$ is clearly found (see \cite{giorgini} and Fig. \ref{fig:susc}).
\bibitem{he3} 
In liquid $^3$He the value fo $F_0^a$ becomes closer and closer to $-1$ as the pressure is increased, revealing a nearly ferromagnetic effect. At very high pressures, however, the liquid becomes a solid before the ferromagnetic transition appears.
\bibitem{forward}
Note that the quasiparticle forward scattering sum rule 
$\sum_l(F_l^s/(1+F_l^s/(2 l+1))+\sum_l(F_l^a/(1+F_l^a/(2 l+1))=0$
is satisfied at the first order in $k_F a$, since only $F_0^{s(a)}$ are affected by interaction at this order.
At the second order all the angular momentum channels contribute to the sum rule.
\end{thebibliography}
\end{document}